%% file: Main.tex
\documentclass[12pt, letterpaper]{article}
\usepackage[margin=2cm]{geometry}

\usepackage{amssymb}
\usepackage{amsmath}
\usepackage{graphicx}
\usepackage{physics}

\usepackage{cite}
\usepackage{url}

\usepackage{color}
\newcommand{\change}[1]{{#1}}

\title{Legendre transformation and information geometry for the maximum entropy theory of ecology}

\author{ Pedro Pessoa$^1$ \\
$^1$Department of Physics, University at Albany - SUNY \\  Albany, NY - USA}
\date{}

\begin{document}
\maketitle
\abstract{

Here I investigate some mathematical aspects of the maximum entropy theory of ecology (METE). In particular I address the geometrical structure of METE endowed by information geometry.
As novel results, the macrostate entropy is calculated analytically by the Legendre transformation of the log-normalizer in METE.  This result allows for the calculation of the metric terms in the information geometry arising from METE and, by consequence, the covariance matrix between METE variables.
}


\newpage
%\tableofcontents
\input{text}

%\section*{Acknowledgments}\paragraph{}

\input{app}
%\newpage

%\bibliographystyle{unsrt}

%\footnotesize
\bibliographystyle{naturemag-doi}
\bibliography{ref}

%\tableofcontents

\end{document}

%% file: text.tex
\section{Introduction}
%\paragraph{}
The method of maximum entropy (MaxEnt) is usually associated with Jaynes'  work \cite{Jaynes57,Jaynes65,Jaynes03} connecting statistical physics and the information entropy proposed by Shannon \cite{Shannon48} -- although its mathematics is known since Gibbs \cite{Gibbs02}. It consists of selecting probability distributions by maximizing a functional -- namely entropy -- usually under a set of expected values constraints, arriving at what is known as Gibbs distributions. Since Shore and Johnson \cite{ShoreJohnson80} MaxEnt has been understood as a general method for inference -- see also \cite{Skilling88,Caticha03,Vanslette17} -- hence it is not surprising that (i) Gibbs distributions are what is known in statistical theory as exponential family -- the only distributions for which sufficient statistics exist (see e.g. \cite{Daum86}), (ii) MaxEnt encompasses the methods of Bayesian statistics \cite{Giffin06}, and (iii) MaxEnt has found successful applications in several fields of science (e.g \cite{Golan08,Bianconi09,CatichaGolan14,Vicente14,Yong16,Delgado-Bonal16,Martino18,Cimini19,Dixit20,Radicchi20,Caldarelli20}).

One of the scientific fields in which MaxEnt has been successfully applied is macroecology. The work of Harte and collaborators \cite{Harte08,Harte11,Harte14,Brummer19,Newman20} presents what is known as the maximum entropy theory of ecology (METE). It consists of finding, through MaxEnt, a joint conditional distribution for the abundance of a species and the metabolic rate of its individuals. From the marginalization and expected values of the MaxEnt distribution, it is possible to obtain 
(i)   the species abundance distribution (Fisher’s log series), 
(ii)  the species-area distribution,
(iii) the distribution for metabolic rates over individuals,
and (iv) the relationship between the metabolic rate of  individuals in a species and that species abundance 
\change{--- for a comprehensive confirmation of METE with experimental data see \cite{Xiao15}.} In a recent article Harte \cite{Harte21} brings forward the need for dynamical models based on MaxEnt, as METE assume the variables to be static\footnote{{It is relevant to say that Jaynes applied dynamical methods based on information theory for nonequilibrium statistical mechanics \cite{Jaynes79} leading to what is now known as maximum caliber  \cite{Presse13,Gonzalez16}. However, maximum caliber assumes a Hamiltonian dynamics and, therefore, does not generalize to ecology and other complex systems.  }}.
.

The field known as information geometry (IG) \cite{Caticha15,Amari16,Ay17,Nielsen20} assigns a Riemannian geometry structure to probability distributions. In information geometry the distances are given by the Fisher-Rao information metric (FRIM) \cite{Fisher25,Rao45}, which is the only metric in accordance with the grouping property of probability distributions \cite{Cencov81}. IG has found important applications for probabilistic dynamical systems \cite{Amari16,Hayashi16,Felice18,Ruppeiner95,Pessoa20}. Here the FRIM terms for the distributions arising from METE will be calculated.
In a future publication I will build upon the results obtained here towards an entropic dynamics \cite{Pessoa20} using METE variables.

The layout of the paper is as follows: The following section (2) presents MaxEnt in general terms followed by the MaxEnt process in METE. In particular we obtain the macrostate entropy through the Legendre transform, and the Lambert W special function \cite{Corless16,Lehtonen16}, which is a novel result to the best of my knowledge. Section 3 presents some general results of IG and calculate the information metric terms for METE. Section 4 concludes the present article by commenting on possible applications and perspectives for IG in a dynamical theory of macroecology.

\section{Maximum Entropy} 
%\paragraph{}

In information theory, probability distributions encode the available information about a system's variables 
$x \in \mathcal{X}$. MaxEnt consists of updating from a prior distribution $q(x)$ -- usually, but not necessarily, taken to be uniform -- to a posterior $\rho(x)$ that maximizes the entropy functional under a set of constraints meant to represent the known information about the system. Usually these constraints are the expected values $A^i$ of a set of real valued functions $\{a^i(x)\}$ \, namely sufficient statistics. The distribution $\rho$ is found as the solution to the following optimization problem

\begin{subequations} \label{maxent}
\begin{align}
\max_\rho \hspace{.5cm}  &  H[\rho] = - \int \dd x \ \rho(x) \log \left( \frac{\rho(x)}{q(x)} \right) \ , \\
s.t. \hspace{.5cm} &  \int  \dd x \ \rho(x) = 1 \\
& \int  \dd x \ a^i(x) \rho(x) = A^i \ .
\end{align} \end{subequations}
where $\int  \dd x$ refers to the appropriate measure of the set $\mathcal{X}$; if one is interested in a discrete set $\mathcal{X} = \{x_\mu\}$, where $\mu$ corresponds to an enumeration of $\mathcal{X}$, we have $\int  \dd x = \sum_\mu$, if one is interested in a continuous subset of real variables, e.g. $\mathcal{X} = [a,b]$,  we have $\int  \dd x = \int_{a}^{b}  \dd x$.

The solution of \eqref{maxent} is the Gibbs distribution

\begin{equation}\label{canonicaldefinition}
    \rho(x|\lambda_1,\lambda_2, ... , \lambda_n) = \frac{q(x)}{Z(\lambda)} \exp \left(-\sum_{i=1}^n \lambda_i a^i(x)\right) \ ,
\end{equation}
where $\lambda = \{\lambda_i\}$ is the set of Lagrange multipliers dual to the expected values $A = \{A^i\}$ 
and $Z(\lambda)$ is a normalization factor given by 

\begin{equation}\label{Z}
    Z(\lambda) = \int  \dd x \ q(x) \exp \left(- \lambda_i a^i(x)\right) \ .
\end{equation}
Above, and on the remainder of this article, we use Einstein's summation notation $A_i B^i = \sum_{i} A_i B^i$.
The expected values can be recovered as

\begin{equation}\label{expected}
    A^i = - \frac{1}{Z} \pdv{Z}{\lambda_i} = \pdv{F}{\lambda_i} \ , \qq{where} F(\lambda) \doteq -\log(Z(\lambda)) \ .
\end{equation}
We will refer to $F$ as the log-normalizer, which displays a role similar to free energy in statistical mechanics.

If one is able to invert the equations arriving from \eqref{expected}, obtaining this way $\lambda_i(A)$ they can express the probability distributions in terms of the expected values, $\rho(x|A) = \rho(x|\lambda(A))$. This also allows one to calculate the entropy $H$ at its maximum -- that means $H[\rho(x|A)]$ for $\rho$ in \eqref{canonicaldefinition} -- as a function of the expected values, rather than a functional of $\rho$, obtaining

\begin{equation}\label{macroentropy}
H(A) \doteq H[\rho(x|\lambda(A))] = 
- \int dx \ \rho(x|\lambda(A)) \log\frac{ \rho(x|\lambda(A))}{q(x)}=  \lambda_i(A) A^i - F (\lambda(A)) \ .
\end{equation}
We will refer to $H(A)$ as the macrostate entropy, which is what we refer to in statistical mechanics as thermodynamical entropy -- meaning the one that appears in the laws of thermodynamics\footnote{Since the arguments that identify the macrostate entropy as the thermodynamical entropy assume that the sufficient statistics are conserved quantities in a Hamiltonian dynamics \cite{Jaynes65},  analogous `laws of thermodynamics' - e.g. conservation of $A^2$ in \eqref{toinvertE} or an impossibility of $H$ in \eqref{meteentropy} to decrease --  are not expected in ecological systems.   }.
.
One can see from \eqref{macroentropy} that $H(A)$ is the Legendre transformation \cite{Nielsen10} of $F(\lambda)$. It also follows that $\lambda_i = \pdv{H}{A^i}$.

%\newpage
\subsection{METE} 
%\paragraph{}
The first step towards a MaxEnt description involves choosing the appropriate variables for the problem at hand. In METE \cite{Harte11} one assumes an ecosystem of $S$ species supporting $N$ individuals with a total metabolic rate $E$, meaning in a unit of time the ecosystem consumes a quantity $E$ of energy. The state of the system $x$ on MaxEnt is defined for a singular species as the number of individuals (abundance) $n$, $n\in \{1,2, \ldots, N\}$ and the metabolic rate of an individual of that species $\varepsilon$, $\varepsilon \in [1,E]$ -- note that one can choose a system of units so that the smallest metabolic rate is the unit, $\varepsilon_{min} = 1$. We represent the state as $x = (n,\varepsilon)$.

The second step consists of assigning the sufficient statistics that appropriately captures the information about the system. In METE \cite{Harte11} the statistics chosen are the number of individuals in the species $a^1(n,\varepsilon) \doteq n$ and the total metabolic rate $a^2(n,\varepsilon) \doteq n\varepsilon$. Substituting  these into the defined expected value constrains for the sufficient statistics \eqref{maxent}, we obtain constraints on average abundance per species

\begin{equation}
    A^1 = \sum_{n=1}^N \ \int_1^E \dd \varepsilon  \ n \ \rho(n,\varepsilon|\lambda) = \frac{N}{S} \doteq N' \ ,
\end{equation}
and a constrain on the average metabolic consumption per species

\begin{equation}
    A^2 = \sum_{n=1}^N \ \int_1^E \dd \varepsilon  \ n \varepsilon \ \rho(n,\varepsilon|\lambda) = \frac{E}{S} \doteq E' \ .
\end{equation}
The defined variable $N'$ and $E'$ will replace $A^1$ and $A^2$, respectively, when convenient.

Having the state variables and the sufficient statistics chosen, we can compute all quantities defined in the previous subsection for the specific system defined by METE. With a uniform prior $q$, this leads to the canonical distribution \eqref{canonicaldefinition} of the form

\begin{equation}\label{metecanonical}
    \rho(n,\varepsilon|\lambda) = \frac{1}{Z(\lambda)} e^{-\lambda_1 n} e^{ -\lambda_2 n\varepsilon} \ ,
\end{equation}
where the normalization factor \eqref{Z} is given by

\begin{equation}\label{metepartition}
    Z(\lambda) = \sum_{n=1}^N \ \int_1^E \dd \varepsilon  \  e^{-\lambda_1 n} e^{ -\lambda_2 n\varepsilon} = \sum_{n=1}^N e^{-\lambda_1 n} \left( \frac{e^{-\lambda_2 n} - e^{-\lambda_2 n E}  }{\lambda_2 \ n} \right) \ ,
\end{equation}
from which the expected values \eqref{expected} can be calculated as

\begin{subequations}\label{meteexpectedvalues}\begin{align}
        A^1 = N' & =   \frac{1}{\lambda_2 \ Z(\lambda)} \sum\limits_{n=1}^N e^{-\lambda_1 n} (e^{-\lambda_2 n} - e^{-\lambda_2 n E} ) \ , \label{Nprime} \\
        A^2 = E' & =  \frac{1}{\lambda_2} \left[ 1 + \frac{1}{Z(\lambda)} \sum\limits_{n=1}^N e^{-\lambda_1 n} (e^{-\lambda_2 n} - E e^{-\lambda_2 n E} ) \right]  \ . \label{Eprime}
\end{align} \end{subequations}
These are complicated equations, however some approximations may make them more treatable. 

A fair assumption, knowing what the variables are supposed to represent, is that there are far more individuals than species, $N\gg S$ and the average metabolic rate per individual is far greater than the unit of metabolic rate $E/N = E'/N' \gg 1$. 
This allows for a sequence of approximation that we will treat like assumptions here, namely 
(i) $ e^{-\lambda_2 n E}\ll e^{-\lambda_2 n} $, 
(ii) $ E e^{-\lambda_2 n E} \ll e^{-\lambda_2 n}$, 
(iii) $\lambda_1+\lambda_2 \ll 1 $, 
and (iv) $e^{- (\lambda_1+\lambda_2 )N} \ll 1$. 
Further explanation on the validity of these assumptions, under $S \ll N \ll E$, can be seen in \cite{Harte11,Brummer19} and their confirmation by numerical calculation can be seen in \cite{Harte11}.
Under this understanding we can substitute \eqref{metepartition} into \eqref{Nprime} obtaining

\begin{subequations}\label{toinvertN}\begin{align}
    N' &= \frac{\sum\limits_{n=1}^N e^{-\lambda_1 n} (e^{-\lambda_2 n} - e^{-\lambda_2 n E} )}{\sum\limits_{n=1}^N \frac{1}{n} e^{-\lambda_1 n} \left( {e^{-\lambda_2 n} - e^{-\lambda_2 n E}  }\right) } \approx \frac{\sum\limits_{n=1}^N e^{-(\lambda_1+\lambda_2) n} }{\sum\limits_{n=1}^N \frac{1}{n} e^{-(\lambda_1+\lambda_2) n}  } \label{notfullyapr}\\ N' &\approx -\left[ \frac{1}{(\lambda_1+\lambda_2) \log({\lambda_1+\lambda_2}) }\right]\ .\label{fullyapr}
\end{align} \end{subequations}
We can also rewrite \eqref{Eprime} obtaining

\begin{equation}\label{toinvertE}
    E' =  \frac{1}{\lambda_2} + \frac{\sum\limits_{n=1}^N e^{-\lambda_1 n} (e^{-\lambda_2 n} - E e^{-\lambda_2 n E} )}{\sum\limits_{n=1}^N \frac{1}{n} e^{-\lambda_1 n} \left( {e^{-\lambda_2 n} - e^{-\lambda_2 n E}  }\right) }  \approx  \frac{1}{\lambda_2} +N' \ .
\end{equation}

In order to obtain the macrostate entropy analytically \eqref{macroentropy}
one needs to perform the Legendre transformation for METE, which includes inverting \eqref{toinvertN} and \eqref{toinvertE} obtaining $\lambda_1(N',E')$ and $\lambda_2(N',E')$. In page 149 of \cite{Harte11} it is said to be unfeasible. However, it is possible to do so obtaining

\begin{equation}\label{inverted}
    \lambda_1 = \beta(N') - \frac{1}{E'-N'} \ , \qq{and} \ \lambda_2 = \frac{1}{E'-N'} \ ,
\end{equation}
where
\begin{equation}\label{betas}
    \beta(N') \doteq - \left[N' \ W_{-1}\left(-\frac{1}{N'}\right)\right]^{-1} \ ,  \qq{}  \dot\beta(N') \doteq \dv{\beta}{N'} = \left[N'^2 - \frac{N'}{\beta(N')}\right]^{-1} %=- \left[N'^2 \left(1+ W_{-1}\left(-\frac{1}{N'}\right) \right)\right]^{-1} 
    \ ,
\end{equation}
and $W_{-1}$ refers to the second main branch of the Lambert W function (see \cite{Corless16,Lehtonen16}). The details on how \eqref{inverted} inverts \eqref{toinvertN} and \eqref{toinvertE} are presented in Appendix \ref{appendixA}. 
The macrostate entropy can be calculated directly from \eqref{macroentropy} as

\begin{equation}\label{meteentropy}
    H(N',E') = N'\beta(N') + \log(E'-N') - \log\left(  {N'\beta(N')}\right)  +1 \ .
\end{equation}
With the calculation of the macrostate entropy finished, we can move into a geometric description of METE.

\section{Information geometry} 
%\paragraph{}
 This section presents the elementary notions of IG -- for more in depth discussion and examples see e.g. \cite{Amari16,Caticha15,Ay17,Nielsen20} -- and some useful identities for the IG of Gibbs distributions. IG consists of assigning a Riemmanian geometry structure to the space of probability distributions, meaning if a set of distributions $p(x|\theta)$ is parametrized by a finite number of coordinates, $\theta = \{ \theta^i\}$, the distances -- which are a measure of distinguishability -- $\dd \ell$ between the neighbouring distributions $P(x|\theta+\dd \theta)$ and $P(x|\theta)$ are given by $\dd \ell^2 = g_{ij} \dd \theta^i \dd \theta^j$. 
 The work of Cencov \cite{Cencov81} demonstrated that the only metric invariant under Markov embeddings -- and, therefore, the only one adequate to represent a space of probability distributions -- is the metric of the form
 
\begin{equation}
\label{firstmetric}
    g_{ij} = \int \dd x \ P(x|\theta) \frac{\partial\log P(x|\theta)}{\partial \theta^i}\frac{\partial\log P(x|\theta)}{\partial \theta^j}  \ ,
\end{equation}
know as FRIM. 

Considering the MaxEnt results presented in previous section, we can restrict our investigation to the Gibbs distributions using the expected values $A$ as coordinates -- $\theta^i = A^i$ and $P(x|\theta) = \rho(x|A)$ as in \eqref{canonicaldefinition}. Two useful expressions arise in that case -- for proofs see e.g. \cite{Caticha15} -- first: the metric terms are the Hessian of the negative of macrostate entropy, meaning

\begin{equation}\label{HessianMetric}
    g_{ij} =- \pdv{H}{A^i}{A^j} = - \pdv{\lambda_i}{A^j}  \ , % \qq{and } g^{ij} =- \pdv{F}{\lambda_i}{\lambda_j}\ ;
\end{equation}
and second:  the covariance matrix between the sufficient statistics $a^i(x)$ is the inverse matrix of $g_{ij}$, meaning

\begin{equation} \label{cov}
    C^{ij}g_{jk} = \delta^i_k \ , \qq{where}C^{ij} = \expval{a^i(x) a^j(x)} - A^iA^j \ .
\end{equation}
We can, then, see how these quantities are calculated for METE.

\subsection{Information geometry of METE} 
%\paragraph{}
By substituting the macrostate entropy for METE \eqref{meteentropy} in \eqref{HessianMetric} we obtain the FRIM terms:

\begin{equation}\label{metemetric}\begin{split}
    g_{11} = - \dot\beta (N') +\frac{1}{(E'-N')^2} \ ,  \quad  g_{12} =&\   g_{21} = - \frac{1}{(E'-N')^2} \ , \\
    g_{22} = \frac{1}{(E'-N')^2} \ , \hfill  \qq{and}  g =& - \frac{\dot\beta (N') }{(E'-N')^2}\ .
\end{split} \end{equation}
where $g = \det g_{ij}$.  Per \eqref{cov} and from the general form of inverse matrix of a two dimensional matrix, the covariance matrix terms can be calculated directly inverting \eqref{metemetric} obtaining

\begin{equation}\label{metecov}\begin{split}
    C^{11} = \frac{g_{22}}{g} = \frac{N'}{\beta (N')} - N'^2 \ ,  
    \quad  C^{12} = C^{21} = - \frac{g_{12}}{g} =  \frac{N'}{\beta (N')} - N'^2  \ ,  \\ 
    \qq{and}    C^{22} = \frac{g_{11}}{g} = E'^2-2 E' N' + \frac{N'}{\beta (N')}  \ \ ; \quad \quad
\end{split} \end{equation}
completing the calculation. The matrix $C^{ij}$ can be interpreted directly as the covariance  between a species abundance and its total metabolic rate -- METE sufficient statistics.  The information metric terms presented in \eqref{metemetric} allow for further studies on dynamical ecology from a information theory background, as we will comment in the following section.

\section{Discussion and perspectives} 
%\paragraph{}
The present article calculates the macrostate entropy \eqref{meteentropy} for METE. This was made possible by the analytical calculation of the Lagrange multipliers \eqref{inverted} as functions of the expected values \eqref{meteexpectedvalues}, previously believed to be unfeasible. 
This allows for a complete description of METE in terms of the average abundance $N'$ and the expected metabolic rate $E'$ of each of the ecosystem species. This opens a broad range of investigations possible by  analytical calculations. In particular, the IG  arising from METE is presented by calculating the FRIM terms in \eqref{metemetric}. Independently of any geometric interpretation, that was equivalent to calculate the covariance between METE sufficient statistics \eqref{metecov}.

The variables that define an ecosystem's state are not expected to remain constant. Because of this, and the growing relevance of IG in dynamical systems, the calculations made in the present article are an important step into expanding maximum entropy ideas into further investigation in macroecology. One possible example, that I intend to explore in a future publication, is using the results presented here towards an entropic dynamics for ecology.

%Lorem ipsum dolor sit amet, consectetur adipiscing elit, sed do eiusmod tempor incididunt ut labore et dolore magna aliqua. Ut enim ad minim veniam, quis nostrud exercitation ullamco laboris nisi ut aliquip ex ea commodo consequat. Duis aute irure dolor in reprehenderit in voluptate velit esse cillum dolore eu fugiat nulla pariatur. Excepteur sint occaecat cupidatat non proident, sunt in culpa qui officia deserunt mollit anim id est laborum

%% file: app.tex
%\newpage
%\section*{Appendix}
\appendix

\section{On the Lambert W function} \label{appendixA}

In this appendix we will explain how \eqref{inverted} inverts \eqref{toinvertN} and \eqref{toinvertE}.
The Lambert W function is defined as the solution of 

\begin{equation}\label{defW}
    W(x) e^{W(x)} = x \ .
\end{equation}
The python library SciPy \cite{scipy} implements the numerical calculation of $W$. This relates to \eqref{fullyapr} in the following manner: by defining the variable $\beta=\lambda_1+\lambda_2$ we obtain

\begin{equation} \label{22}
    \frac{1}{N'} = -\beta \log \beta \iff \frac{1}{\beta N'} e^{-\frac{1}{\beta N'}} = \frac{1}{N'} \ ,
\end{equation}
hence $\beta = - \left[N' W\left(-\frac{1}{N'}\right)\right]^{-1}$. It is relevant to say that, from \eqref{defW}, $W(x)$ is  multivalued -- the terminology Lambert W `function' is used loosely. The several single-valued functions that solve \eqref{defW} are known as the different `branches' of the Lambert W. In \eqref{inverted} and \eqref{betas} only the $W_{-1}$ branch was taken into account. Given our object of study, we will restrict to functions that are guaranteed to give a $\beta$ that is real for large $N'$. As explained in \cite{Corless16}, the two branches $W_0(x)$ and $W_{-1}(x)$ are real and analytic for $-e^{-1}<x<0$, of equivalently $\beta$ is real for $N'>e$. Coherent with the fact that \eqref{toinvertN} was derived for large $N'$.
\begin{figure}[h]
    \centering
    \includegraphics[width=.5\textwidth]{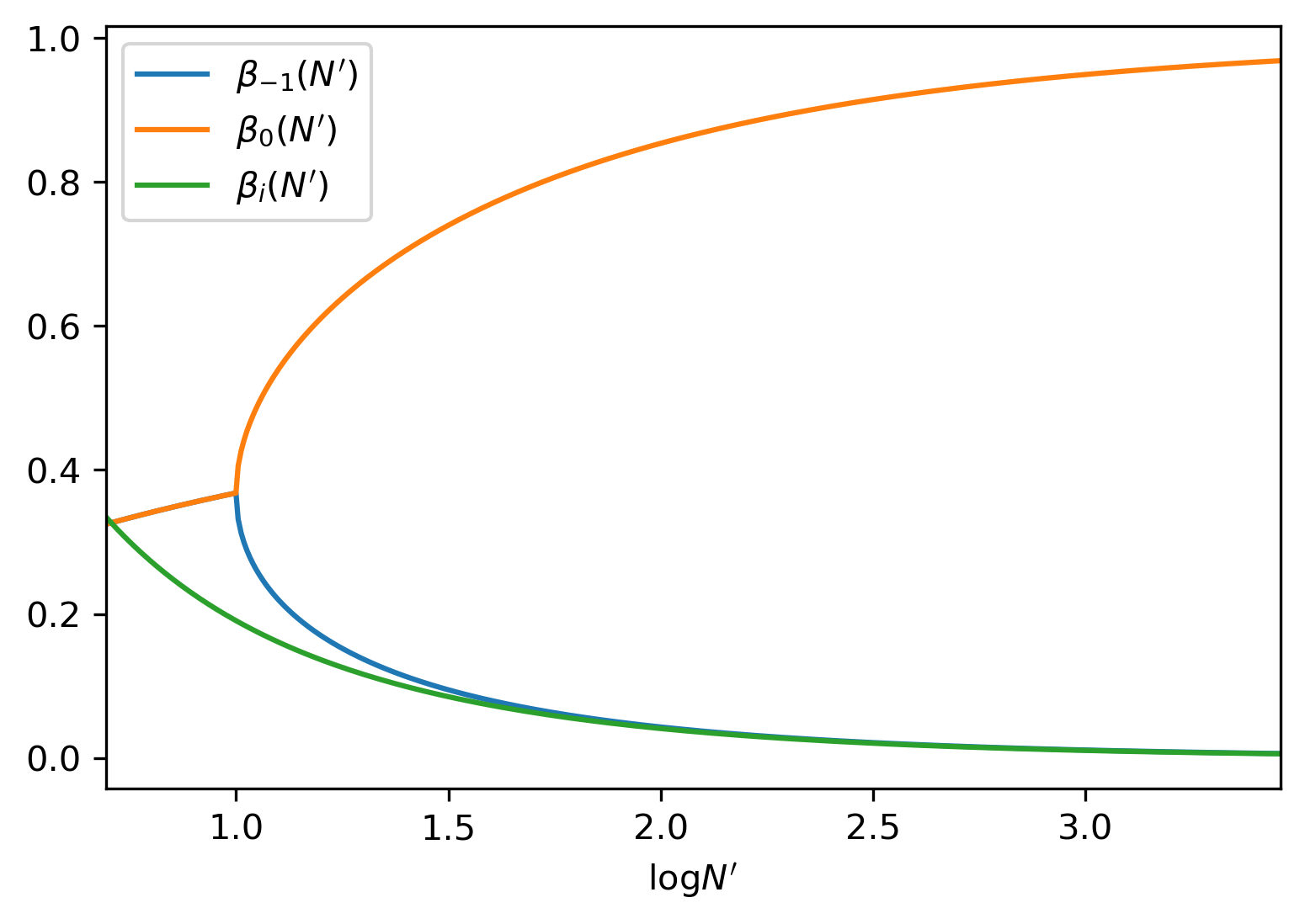}
    \caption{Graphical comparison between the functions defined as: $\beta_0(N') \doteq - \left[N' W_{0}\left(-\frac{1}{N'}\right)\right]^{-1}$, $\beta_{-1}(N')\doteq - \left[N' W_{-1}\left(-\frac{1}{N'}\right)\right]^{-1}$, and $\beta_i(N')$ -- obtained numerically from inverting \eqref{notfullyapr}, here using $S=N/N' = 20$. $W_{0}$ and $W_{-1}$ have complex values for $N'<e$, the graph above only plots the real part in that region. }
    \label{LT}
\end{figure}

%A graph for these the two branches is presented in Fig.
Fig. \ref{LT} presents the graphs of $\beta$ obtained from the $W_0(x)$ and $W_{-1}(x)$ branches, as well as a comparison to the $\beta$ obtained numerically from inverting \eqref{notfullyapr}. Even though per \eqref{22} the $\beta$ obtained by both branches inverts \eqref{fullyapr}, it can be seen from Fig. \ref{LT} that only the  one obtained from $W_{-1}(x)$ approximates the inverse of \eqref{notfullyapr} for large $N'$ and, therefore, it is the only one appropriate for the present investigation.

To complete the claim that $\lambda_1$ and $\lambda_2$ in \eqref{inverted} are calculated analytically, it is relevant  to say that $W_{-1}\left(-\frac{1}{N'}\right)$ can be calculated using the series expansion (see page 153 in \cite{Corless16})

\begin{equation}
    W_{-1}\left(-\frac{1}{N'}\right) = \change{-}\sum_{m=0}^\infty a_m z^m \ , \qq{where} z = \sqrt{2(\log N'-1)} \ ,
\end{equation}
and $a_m$ is defined recursively as $a_0 = 1$, $a_1=1$, and

\begin{equation}
    a_m = \frac{1}{m+1} \left(a_{m-1} -\change{ \sum_{k=2}^{m-1} } k \ a_k \  a_{m+1-k}  \right) \ .
\end{equation}
Note that real $z$ implies $N'>e$, which is coherent with the condition for $W_{-1}$ to be real.
%\cite{Xiao15}